\documentclass[12pt]{iopart}
\usepackage{graphicx,color}
\usepackage{iopams}

\begin{document}

\title[Variational ground states of the 2D Hubbard model]{Variational ground states of the two-dimensional Hubbard model}

\author{D Baeriswyl, D Eichenberger and M Menteshashvili}

\address{Department of Physics, University of Fribourg, 
CH-1700 Fribourg, Switzerland.}

\ead{dionys.baeriswyl@unifr.ch}

\begin{abstract}

Recent refinements of analytical and numerical methods have improved 
our understanding of the ground-state phase diagram of the two-dimensional
(2D) Hubbard model. Here we focus on variational approaches, but comparisons
with both Quantum Cluster and Gaussian Monte Carlo methods are also made. Our
own ansatz leads to an antiferromagnetic ground state at half filling with a
slightly reduced staggered order parameter (as compared to simple mean-field
theory). Away from half filling, we find $d$-wave superconductivity, but
confined to densities where the Fermi surface passes through the
antiferromagnetic zone boundary (if hopping between both nearest-neighbour and
next-nearest-neighbour sites is considered). Our results agree surprisingly
well with recent numerical studies using the Quantum Cluster method. An
interesting trend is found by comparing gap parameters $\Delta$
(antiferromagnetic or superconducting) obtained with different variational
wave functions. $\Delta$ varies by an order of magnitude and thus cannot be
taken as a characteristic energy scale. In contrast, the order parameter is
much less sensitive to the degree of sophistication of the variational
schemes, at least at and near half filling.

\end{abstract}


\submitto{\NJP}

\maketitle

\section{Introduction}
\label{sec:intro}

The conventional mechanism of Cooper pairing rests on phonon exchange, 
but the case of superfluid $^3$He shows that no additional degrees of 
freedom are necessary for establishing pairing, even in a system of fermions
with predominantly repulsive interactions. Several years before the discovery
of superfluidity in $^3$He Kohn and Luttinger used perturbation theory to show 
that even a purely repulsive bare interaction can lead to an effective 
attraction in a channel with large enough angular momentum \cite{Kohn1}. 
An analogous situation may prevail in superconducting
cuprates, as suggested by Anderson as early as 1987 \cite{Ande1}. The Hubbard
Hamiltonian with a large on-site repulsion $U$ is indeed the natural model
for describing the antiferromagnetic Mott insulator of undoped cuprates.
Clearly there are phonons in these materials and the electron-phonon coupling
can be rather strong, as suggested by optical spectroscopy \cite{Mish1}, but
a thorough treatment of both strong correlations and electron-phonon
coupling is required to understand to what extent phonons affect the 
electronic properties in the cuprates. A recent debate \cite{Gius1, Rezn1} 
on the origin of a kink observed in photoemission experiments demonstrates
that such an analysis is still lacking. Correspondingly, the role of phonons
in the superconductivity of cuprates remains also unclear.

In this paper we restrict ourselves to the two-dimensional (2D) Hubbard model,
described by the Hamiltonian
\begin{equation}
\hat{H}=\hat{H}_0+U\hat{D}\, ,
\end{equation}
where
\begin{equation}
\hat{H}_0=-\sum_{ij\sigma}t_{ij}c_{i\sigma}^\dag c_{j\sigma}
\end{equation}
describes hopping over the sites of a square lattice and
\begin{equation}
\hat{D}=\sum_{i}n_{i\uparrow}n_{i\downarrow}
\end{equation} 
measures the number of doubly occupied sites. The operator $c_{i\sigma}$
($c_{i\sigma}^\dag$) annihilates (creates) an electron at site $i$ with
spin $\sigma$ and $n_{i\sigma}=c_{i\sigma}^\dag c_{i\sigma}$. Only hopping
between nearest ($t_{ij}=t$) and next-nearest neighbours ($t_{ij}=t'$) will be
considered.

For small $U$ the approach of Kohn and Luttinger looks promising, but naive
perturbation theory diverges and one has to sum entire diagram classes (see
\cite{Binz2} for a pedagogical treatment of this problem). Moreover, there are
several competing instabilities close to half filling, in particular, $d$-wave
superconductivity and antiferromagnetism, or rather spin-density waves. The
problem can be solved in an elegant way using the functional renormalization
group \cite{Zanc1, Halb1, Hone3, Binz1}, which treats the competing
density-wave and superconducting instabilities on the same footing. The
specific techniques used by the different groups to calculate effective
vertices and susceptibilities differ in details, but the overall results are
more or less consistent, with an antiferromagnetic instability at half filling
and $d$-wave superconductivity away -- but not too far away -- from half
filling.

In the large-$U$ limit, where double occupancy is suppressed, the Hubbard 
model can be replaced by the Heisenberg
model at half filling \cite{Ande3} and by the $t-J$ model close to half filling
\cite{Chao1}, defined by the Hamiltonian
\begin{equation}
\hat{H}_{\rm t-J}
=-\sum_{ij\sigma}t_{ij}\hat{P}_0c_{i\sigma}^\dag c_{j\sigma}\hat{P}_0
+\sum_{ij}J_{ij}\left({\bi S}_i\cdot{\bi S}_j-\frac{1}{4}n_in_j\right)\, ,
\end{equation}
where $n_i=n_{i\uparrow}+n_{i\downarrow}$ counts 
the number of electrons at site $i$, the projector 
\begin{equation}
\hat{P}_0=\prod_i(1-n_{i\uparrow}n_{i\downarrow})
\end{equation}
eliminates configurations with doubly occupied sites, ${\bi S}_i$ are spin 
$\frac{1}{2}$ operators and $J_{ij}=2t_{ij}^2/U$.
 
While the ground state of the 2D Heisenberg model is fairly well understood 
\cite{Manu1} -- it is widely accepted that it exhibits long-range 
antiferromagnetic order with a reduced moment due to quantum fluctuations --
the ground state of the $t-J$ model remains an unsolved problem, despite an
extensive use of sophisticated methods during the last two decades 
\cite{Lee1, Ogat1}. A simple ansatz 
\begin{equation}
\vert\Psi\rangle = \hat{P}_0\vert\Psi_0\rangle\, ,
\label{eq:vanilla}
\end{equation}
is frequently used, where $\vert\Psi_0\rangle$ is the ground state of a 
suitable single-particle Hamiltonian, representing for instance the filled 
Fermi sea, a spin-density wave or a superconductor. In a large region of 
doping the most favourable mean-field state $\vert\Psi_0\rangle$ has been 
found to be a superconducting ground state with $d$-wave symmetry
\cite{Gros1, Yoko1, Para1, Path1}. In this case the ansatz (\ref{eq:vanilla}) 
describes a resonating valence bond (RVB) \cite{Paul1} state, jokingly
referred to as the `plain vanilla version of RVB' \cite{Ande5}. 

Sometimes the $t-J$ model is
used to provide a simple argument for pairing. If the exchange coupling $J$
is larger than the hopping amplitude $t$ then it is favourable for two 
holes to remain nearest neighbours. Unfortunately, this is an unphysical limit
for the Hubbard model for which the $t-J$ model is only a valid approximation
for $J\ll t$.

Experiments on layered cuprates indicate that $U$ is neither small enough
for a perturbative treatment nor large enough for the mapping to the $t-J$ 
model. Thus, neutron scattering experiments on undoped cuprates \cite{Cold1} 
can be very well interpreted in terms of the 2D Hubbard model with $U$ between 
$8\, t$ and $10\, t$ (depending on the approximation used for the magnon 
spectrum \cite{Kata1, Muel1}), in agreement both with an analysis of Raman data
\cite{Kata2} and with a comparison between theoretical and experimental
optical absorption spectra \cite{Coma1}. As to the `kinetic' term 
$\hat{H}_0$, angular-resolved photoemission 
experiments \cite{Aebi1, Bori1} give evidence for a hole-like Fermi surface,
which can be described by taking both nearest-neighbour and 
next-nearest-neighbour hopping terms into account \cite{Norm1, Kim1}; 
the experiments are consistent with $t'\approx-0.3t$.

A lot of effort has been spent to treat the Hubbard model in the crossover 
regime of moderately large $U$. Quantum Monte Carlo methods have been
developed in the 1980s, but they suffer from severe statistical 
uncertainties at low temperatures and for large system sizes, both for
many-fermion and quantum spin Hamiltonians \cite{Dago1}. This 
`minus sign problem' turns out to be intrinsically hard \cite{Troy1}. 
The recently introduced Gaussian Quantum Monte Carlo method \cite{Corn1}
leads to positive weights, but the method is not immune to systematic errors 
\cite{Corb1}. Quantum Cluster methods, which replace the single site of the
Dynamical Mean-Field Theory \cite{Geor1} by a cluster of several sites, 
offer a promising alternative route for the intermediate-$U$ regime 
\cite{Maie2, Sene1}. However, the spatial extent of the cluster that is 
treated essentially exactly has so far been very small ($2\times 2$). The 
large changes found in the case of the Mott transition when proceeding from 
a single site to a ($2\times 2$) cluster \cite{Park1} indicate that
larger cluster sizes are needed to obtain accurate results.

Alongside Quantum Monte Carlo and Quantum Cluster methods, variational schemes
have played a major role in the search of possible 
ground states of many-body systems. In the context of the Hubbard model, many
variational wave functions go back to the ansatz of Gutzwiller \cite{Gutz1}
\begin{equation}
\vert\Psi_{\rm G}\rangle=\rme^{-g\hat{D}}\vert\Psi_0\rangle\, ,
\label{eq:G}
\end{equation}
where the variational parameter $g$ reduces double occupancy, weakly for small 
$U$ and completely for $U\rightarrow\infty$. 
In spite of the simplicity of the ansatz, a substantial
part of the correlation energy is obtained for small values of $U$.
At half filling, the variational ground-state
energy tends to the exact limiting value 0 for $U\rightarrow\infty$, where
double occupancy is suppressed ($g\rightarrow\infty$) and the ansatz
(\ref{eq:G}) coincides with the RVB state 
(\ref{eq:vanilla}). Nonetheless many properties such as spin-spin
correlation functions are described rather poorly, especially in the large-$U$ 
regime. Several improvements have been proposed in terms of additional
operators in front of the Gutzwiller ansatz, either to enhance magnetic or 
charge correlations \cite{Copp1, Yoko4} or to increase the tendency
of empty and doubly occupied sites to be next to each other 
\cite{Kapl1, Yoko2}. Very recently more elaborate states have been proposed,
one including backflow \cite{Tocc1}, the other introducing a large number
of variational parameters, both into $\vert\Psi_0\rangle$ and into the
(Gutzwiller-Jastrow) correlation factor \cite{Taha1}.

The main purpose of this paper is to describe our own variational ansatz in 
some more detail than previously \cite{Eich2, Eich3} and to compare our results
with those of other approaches. In \sref{sec:var} our ansatz is defined
and shown to recover Anderson's RVB  state in the large-$U$ limit and for
small hole doping. \Sref{sec:sq} analyses various wave functions for
the Hubbard model on a four-site plaquette.  For the square lattice, a
variational Monte Carlo method is used to study our ansatz. The procedure is
explained in \sref{sec:AF} and applied to an antiferromagnetic state at
half filling. Our results for superconductivity away from half filling are
presented in \sref{sec:SC} and compared to those of other variational wave
functions and also to a very recent computation using a Quantum Cluster
method. We do find evidence for a superconducting ground state, with an
interesting hint at a magnetic mechanism. The concluding
\sref{sec:sum} gives a brief summary of our main results.

\section{Variational ground states}
\label{sec:var}

The primary effect of the Gutzwiller variational state 
(\ref{eq:G}) is the increased suppression of double occupancy
as a function of $U$. Alternatively, we may motivate the ansatz as follows.
Consider a general Hamiltonian 
$\hat{H}=\hat{H}_0+\hat{H}_{\rm int}$, where $\hat{H}_0$ is a
single-particle operator and $\hat{H}_{\rm int}$ the
interaction term. The state
\begin{equation}
\vert\Psi_\lambda\rangle=\rme^{-\lambda\hat{H}}\vert\Psi_{\rm t}\rangle
\end{equation}
tends to the exact ground state $\vert\Psi\rangle$ of $\hat{H}$ for 
$\lambda\rightarrow\infty$ (or to one of the ground states in the case
of degeneracy) unless the trial state $\vert\Psi_{\rm t}\rangle$ is orthogonal
to $\vert\Psi\rangle$. If $\vert\Psi_{\rm t}\rangle$
does not differ too much from $\vert\Psi\rangle$, a small parameter $\lambda$
may be sufficient to obtain a good approximation.
In this case we can make the replacement
\begin{equation}
\rme^{-\lambda\hat{H}}\approx 
\rme^{-\lambda\hat{H}_{\rm int}}\, \rme^{-\lambda\hat{H}_0}\, .
\label{eq:Factorization1}
\end{equation} 
Choosing $\vert\Psi_{\rm t}\rangle =\vert\Psi_0\rangle$, the ground state of 
$\hat{H}_0$, and treating $\lambda$ as a variational parameter, we recover
the Gutzwiller ansatz (\ref{eq:G}) in the case of the Hubbard model. 
At the same time this derivation indicates that the Gutzwiller ansatz is 
best suited for small values of $U$. 

The factorization (\ref{eq:Factorization1}) is not unique, we may 
also choose
\begin{equation}
\rme^{-\lambda\hat{H}}\approx 
\rme^{-\lambda\hat{H}_0}\, \rme^{-\lambda\hat{H}_{\rm int}}\, .
\label{eq:Factorization2}
\end{equation}
With $\vert\Psi_{\rm t}\rangle=\vert\Psi_0\rangle$ we arrive at an ansatz where
both exponentials have to be kept. Allowing them to vary independently, we 
obtain
\begin{equation} 
\label{eq:GB}
\vert\Psi_{\rm GB}\rangle=\rme^{-h\hat{H}_0/t}\,
\rme^{-g\hat D}\vert\Psi_{0}\rangle\, .
\end{equation}
The operator $\rme^{-g\hat D}$ partially suppresses double occupancy for $g>0$, 
while $\rme^{-h\hat{H}_0/t}$ promotes both hole motion and kinetic exchange. 
The limit $h\rightarrow 0$ leads back to the Gutzwiller ansatz. The
variational state (\ref{eq:GB}) has been introduced by Otsuka
\cite{Otsu1} and studied both in one and in two dimensions. He found a 
substantial improvement with respect to the Gutzwiller ansatz. Moreover, 
$\vert\Psi_{\rm GB}\rangle$ has a large overlap with the exact ground state for
all values of $U$ in one dimension, whereas on a square lattice  this is only 
true for relatively small values of $U$ (smaller than the bandwidth).

For $g\rightarrow\infty$ and $h\ll 1$ the ansatz (\ref{eq:GB})
leads to the $t-J$ model. To see this, we consider the limit 
$U\rightarrow\infty$, where $\rme^{-g\hat{D}}\vert\Psi_0\rangle$
is replaced by the r.h.s.\ of (\ref{eq:vanilla}).
The expectation value of the energy is then given by
\begin{equation}
E=\frac{\langle\Psi_0\vert\hat{P}_0\rme^{-h\hat{H}_0/t}(\hat{H}_0+U\hat{D})
\rme^{-h\hat{H}_0/t}\hat{P}_0\vert\Psi_0\rangle}
{\langle\Psi_0\vert\hat{P}_0\rme^{-2h\hat{H}_0/t}\hat{P}_0\vert\Psi_0\rangle}\, .
\label{eq:energy}
\end{equation}
At half filling the parameter $h$ is equal to $-t/U$ in the large-$U$ limit
\cite{Baer3} and vanishes for $U\rightarrow\infty$.
This remains valid very close to half filling, as we readily demonstrate. 
Expanding the expression (\ref{eq:energy}) in powers of $h$, we obtain
for the numerator
\begin{eqnarray}
\fl\langle\Psi_0\vert\hat{P}_0\rme^{-h\hat{H}_0/t}(\hat{H}_0+U\hat{D})
\rme^{-h\hat{H}_0/t}\hat{P}_0\vert\Psi_0\rangle=&\langle\Psi_0\vert\hat{P}_0\hat{H}_0\hat{P}_0\vert\Psi_0\rangle-2\frac{h}{t}\langle\Psi_0\vert\hat{P}_0\hat{H}_0^2\hat{P}_0
\vert\Psi_0\rangle+\nonumber\\ &+U\frac{h^2}{t^2}\langle\Psi_0\vert\hat{P}_0\hat{H}_0\hat{P}_1\hat{H}_0\hat{P}_0\vert\Psi_0\rangle+...\, ,
\end{eqnarray}
where $\hat{P}_1$ is a projector onto the subspace with one doubly occupied
site.
For small doping the operator $\hat{H}_0^2$ in the second term can be replaced
by $\hat{H}_0\hat{P}_1\hat{H}_0$, and the denominator in (\ref{eq:energy}) by 1. The minimization
of the energy with respect to $h$ then gives $h=-t/U$ and
\begin{equation}
E_{\rm min}\approx
\langle\Psi_0\vert\hat{P}_0
\left[\hat{H}_0-\frac{1}{U}\hat{H}_0\hat{P}_1\hat{H}_0\right]
\hat{P}_0\vert\Psi_0\rangle\, .
\end{equation}
Close to half filling, the second term in this expression can be rewritten as
\begin{equation}
\hat{P}_0\hat{H}_0\hat{P}_1\hat{H}_0\hat{P}_0\approx
2\sum_{i,j}t_{ij}^2\left(\frac{1}{4}n_in_j-{\bi S}_i\cdot{\bi S}_j\right)\, .
\end{equation}
Thus we find indeed
that in the limit $U\rightarrow\infty$ the variational treatment of the
Hubbard model using the ansatz (\ref{eq:GB}) is equivalent to 
the variational treatment of the $t-J$ model using the ansatz 
(\ref{eq:vanilla}). However, we have to keep in mind that the use of (\ref{eq:GB}) for $g\rightarrow\infty$ is of doubtful validity.
Therefore variational results obtained with the fully projected Gutzwiller
wave function, as in the `plain vanilla' RVB theory, have to be handled 
with care.

A complementary approach \cite{Baer3} starts from the ground state 
$\vert\Psi_\infty\rangle$ for $U\rightarrow\infty$, ideally the ground state
of the $t-J$ model close to half filling. Applying the same arguments as above,
we obtain two new variational states,
\begin{eqnarray}
\vert\Psi_{\rm B}\rangle&=\rme^{-h\hat{H}_0/t}\, \vert\Psi_\infty\rangle\,
,\nonumber\\ \vert\Psi_{\rm BG}\rangle&=\rme^{-g\hat{D}}\,\rme^{-h\hat{H}_0/t}\, 
\vert\Psi_\infty\rangle\, ,
\label{eq:B-BG}
\end{eqnarray}
which should be well suited for describing the large-$U$ regime. 
Unfortunately, it is in general very difficult to deal with the variational
states $\vert\Psi_{\rm B}\rangle$ and $\vert\Psi_{\rm BG}\rangle$, because even
an approximate calculation of $\vert\Psi_\infty\rangle$ represents already
a highly nontrivial problem \cite{Sore1}.

The trial states (\ref{eq:G}), (\ref{eq:GB}) and
(\ref{eq:B-BG}) yield an appealing picture for the Mott metal-insulator
transition as a function of $U$ at half filling \cite{Dzie1, Dzie2, Mart1,
  Baer4}, where both $\vert\Psi_{\rm G}\rangle$ and $\vert\Psi_{\rm GB}\rangle$
represent metallic states, while $\vert\Psi_{\rm B}\rangle$ and
$\vert\Psi_{\rm BG}\rangle$ are insulating. For the soluble 1D Hubbard model with
long-range hopping, where  $t_{ij}\sim \vert i-j\vert^{-1}$, the variational
energies for $\vert\Psi_{\rm G}\rangle$ and $\vert\Psi_{\rm B}\rangle$ are
dual to each other at half filling; similarly $\vert\Psi_{\rm GB}\rangle$ and
$\vert\Psi_{\rm BG}\rangle$ form a dual pair \cite{Dzie1}. On the metallic side
($U$ smaller than the bandwidth) $\vert\Psi_{\rm GB}\rangle$ yields an order of
magnitude improvement for the ground state energy as compared to
$\vert\Psi_{\rm G}\rangle$, and the same effect is observed on the insulating
side when $\vert\Psi_{\rm B}\rangle$ is replaced by $\vert\Psi_{\rm BG}\rangle$.

\section{Lessons from the Hubbard square}
\label{sec:sq}

\begin{figure}
\centering
\includegraphics[width=0.9\textwidth]{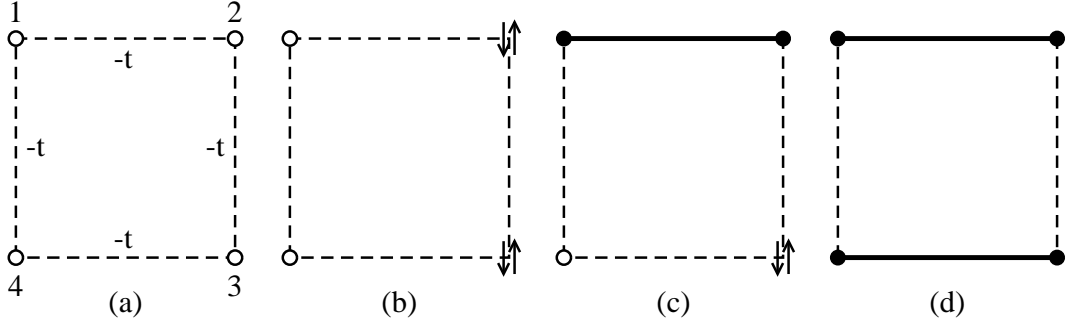}
\caption{(a) Illustration of the Hubbard square\,; (b)-(d) representations of
  the three different types of configurations present in the ground state of
  the half-filled Hubbard square. The symbols stand for: $\opencircle$ empty
  site, $\fullcircle\!\!\!\full\!\fullcircle$ singlet bond,
  $\boldsymbol{\uparrow}\!\boldsymbol{\downarrow}$ doubly  occupied site.}
\label{fig:sq}
\end{figure}

The study of small clusters can provide useful benchmarks for variational
wave functions. Clearly the cluster size should not be too small. In fact, for
two particles on two sites both $\vert\Psi_{\rm G}\rangle$ and $\vert\Psi_{\rm
  B}\rangle$ yield the exact ground state, and there is no need to use more
elaborate wave functions. Therefore we proceed to the Hubbard square, an
essentially 1D system when hopping is restricted to pairs of adjacent
sites (\fref{fig:sq}(a)). We consider first the case of four particles and
denote the corresponding ground state by $\vert\Psi^{(4)}\rangle$. A theorem
by Lieb \cite{Lieb1} directly implies that the ground state of the 1D Hubbard
model for $N$ particles on $N$ sites is non-degenerate for any  positive value
of $U$ and has total spin $S=0$. For $U=0$ the ground state of our Hubbard
square is fourfold degenerate (for $S_z=0$) with energy  $E=-4\, t$, and one
can use perturbation theory to determine the $U\rightarrow 0$ limit of
$\vert\Psi^{(4)}\rangle$. Apart from $S=0$, $\vert\Psi^{(4)}\rangle$ can be
characterized by its transformation properties with respect to the symmetry
group ${\cal D}_4$; it has total quasimomentum $\pi$ ({\it i.e.} it is odd
under the rotation of the square by the angle $\pi/2$ about the axis
perpendicular to the plane of the square and passing through its centre), it
is even with respect to reflections in a plane parallel to one of the sides
and odd with respect to reflections in a plane passing through a diagonal.
Denoting $d^{\dag}_{i}:= c^{\dag}_{i\uparrow}c^{\dag}_{i\downarrow}$ and 
$b^{\dag}_{ij}:=\frac{1}{\sqrt2}(c^{\dag}_{i\uparrow}c^{\dag}_{j\downarrow}
-c^{\dag}_{i\downarrow}c^{\dag}_{j\uparrow})=b^{\dag}_{ji}$,
the ground state can be written as
\begin{equation}
\label{eq:exact}
\vert\Psi^{(4)}\rangle=\frac{a\vert\Phi_{2}\rangle
+b\vert\Phi_{1}\rangle+c\vert\Phi_{0}\rangle}{\sqrt{a^{2}+b^{2}+c^{2}}},
\end{equation} 
with
\begin{eqnarray}
\label{eq:basis}
\fl\vert\Phi_{2}\rangle&=\frac{1}{2}(d^{\dag}_{1}d^{\dag}_{2}
-d^{\dag}_{2}d^{\dag}_{3}+d^{\dag}_{3}d^{\dag}_{4}-d^{\dag}_{1}d^{\dag}_{4})
\vert0\rangle,\nonumber\\
\fl\vert\Phi_{1}\rangle&=\frac{1}{2\sqrt{2}}(d^{\dag}_{1}b^{\dag}_{23}
-d^{\dag}_{2}b^{\dag}_{34}+d^{\dag}_{3}b^{\dag}_{41}-d^{\dag}_{4}b^{\dag}_{12}
-d^{\dag}_{1}b^{\dag}_{43}+d^{\dag}_{2}b^{\dag}_{14}
-d^{\dag}_{3}b^{\dag}_{21}+d^{\dag}_{4}b^{\dag}_{32})\vert0\rangle,\nonumber\\
\fl\vert\Phi_{0}\rangle&=\frac{1}{\sqrt3}(b^{\dag}_{14}b^{\dag}_{23}
-b^{\dag}_{12}b^{\dag}_{34})\vert0\rangle,\nonumber
\end{eqnarray}
where $\vert0\rangle$ is the vacuum (zero-electron) state and $a,b,c$
are real, $U$-dependent coefficients, whose analytical expressions 
are too cumbersome to be presented here. The states 
$\vert\Phi_D\rangle$, illustrated in \fref{fig:sq}(b)-(d), are eigenstates of $\hat{D}$ with $D=0,1,2$ doubly 
occupied sites (doublons). With increasing $U$ the weights of 
$\vert\Phi_1\rangle$ and $\vert\Phi_2\rangle$ decrease and vanish for 
$U\rightarrow\infty$. Therefore $\vert\Phi_{0}\rangle$ is equal to 
$\vert\Psi_\infty\rangle$, the normalized ground state of the Heisenberg 
model on the square with nearest-neighbour antiferromagnetic exchange. We 
also note that in the state $\vert\Phi_{2}\rangle$ the two doublons sit on 
adjacent sites of the square. Similarly, in the state $\vert\Phi_{1}\rangle$ 
the doublon and the holon (empty site) sit on adjacent sites of the 
square. The hopping operator $\hat{H}_0$ mixes the states $\vert\Phi_D\rangle$,
but does not bring in other states. 

As already noted by other authors \cite{Scal3}, the ground state 
$\vert\Psi^{(4)}\rangle$ has an affinity for $d$-wave pairing. In fact,
the quantity $\langle\Psi^{(4)}\vert C_d C^\dag_d\vert\Psi^{(4)}\rangle$,
where $C^\dag_d:=\frac{1}{2}(b_{12}^\dag-b_{23}^\dag+b_{34}^\dag-b_{41}^\dag)$
creates a $d$-wave pair, is larger than the corresponding quantity for 
$s$-wave pairing,
$\langle\Psi^{(4)}\vert C_s C^\dag_s\vert\Psi^{(4)}\rangle$ with
$C^\dag_s:=\frac{1}{2}(b_{12}^\dag+b_{23}^\dag+b_{34}^\dag+b_{41}^\dag)$.
The difference, especially prominent for small $U$, decreases with increasing
$U$ and tends to zero for $U\rightarrow\infty$. The ground state 
$\vert\Psi^{(4)}\rangle$ also exhibits local antiferromagnetic order, which
is noticeably enhanced with increasing $U$.

\begin{figure}
\includegraphics[width=0.49\textwidth]{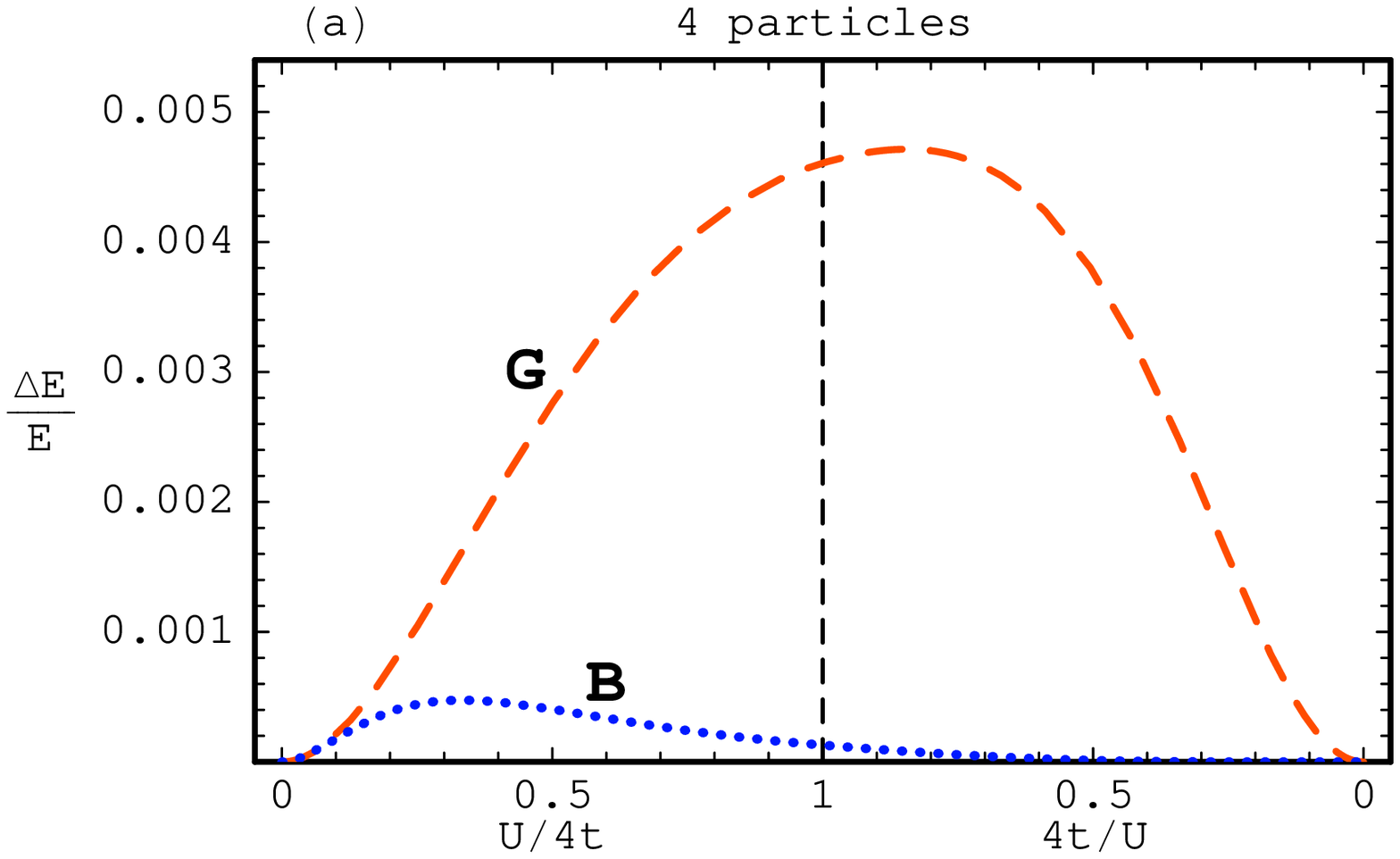}
\hfill
\includegraphics[width=0.49\textwidth]{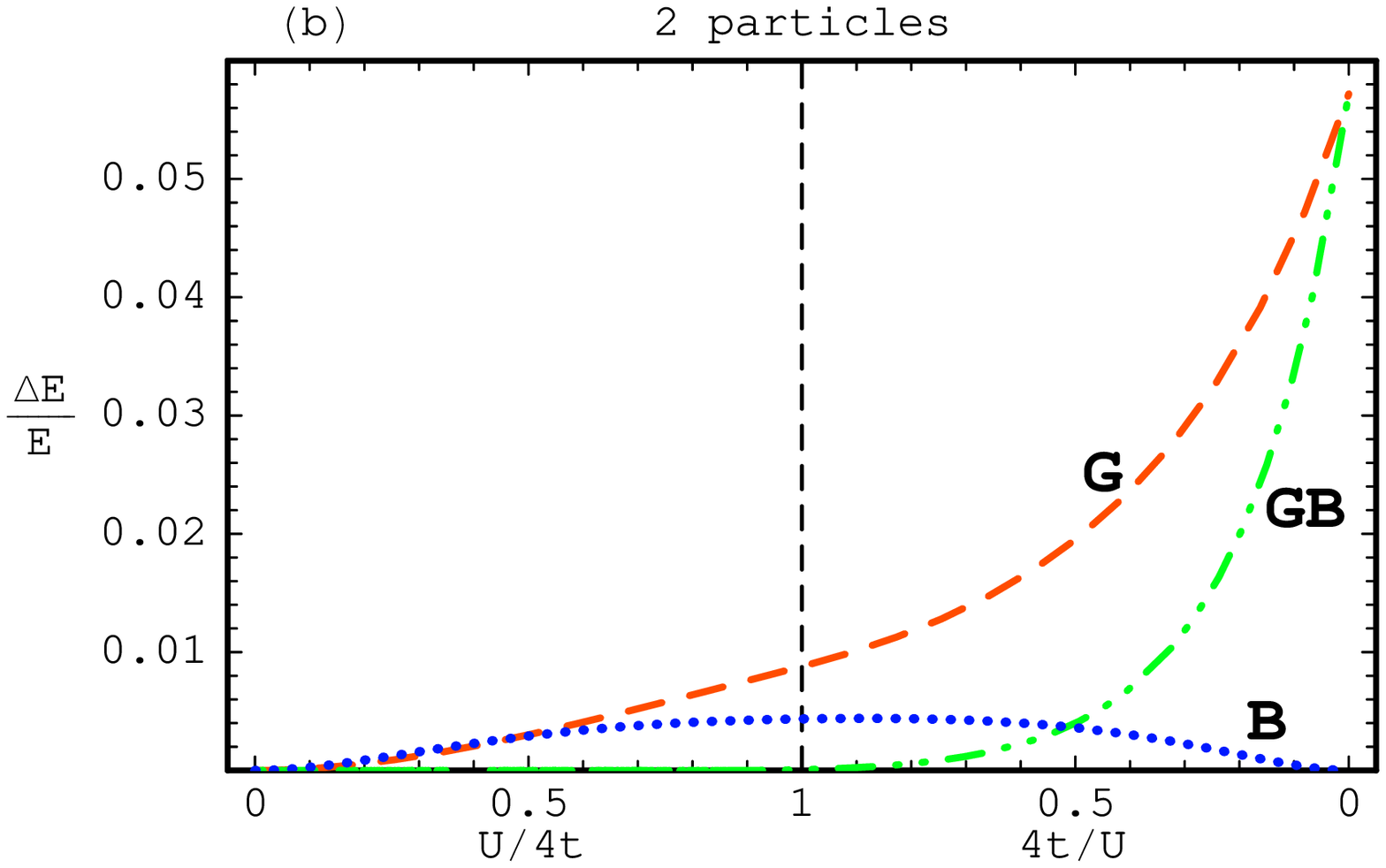}
\caption{Relative deviation $\Delta E/E$ of different variational ground-state
 energies from the exact value $E$ for the Hubbard square:\,(a) half filling,
 (b) quarter filling. The lines correspond to:
${\color{red}\broken}$\quad$\vert\Psi_{\rm G}\rangle$,
 ${\color{blue}\dotted}$\quad$\vert\Psi_{\rm B}\rangle$,\,
 ${\color{green}\dashddot}$\quad$\vert\Psi_{\rm GB}\rangle$.}
\label{fig:DeltaE/E}
\end{figure}

Knowing the exact ground state, one can examine the quality of various
variational wave functions. \Fref{fig:DeltaE/E}(a) shows that the Gutzwiller ansatz $\vert\Psi_{\rm G}\rangle$
already yields a fairly good approximation for all values of $U$, with the
biggest deviation for intermediate $U$ ($\sim 4\, t$), where the variational
energy deviates from the exact value by about $0.5\%$, whereas the overlap 
$\langle\Psi_{\rm G}\vert\Psi^{(4)}\rangle$ is still about $99.95\%$. The state 
$\vert\Psi_{\rm B}\rangle$, equation (\ref{eq:B-BG}), is even better, with an 
improvement of one order of magnitude (or more) for $U>2\, t$ as compared to 
$\vert\Psi_{\rm G}\rangle$.
Generally speaking, since the wave function (\ref{eq:exact}) has two 
independent coefficients, one needs in principle a variational wave function 
with at least two parameters to reproduce the exact ground state.
Not all wave functions are equally successful, though. Both the refined 
Gutzwiller ansatz $\vert\Psi_{\rm GB}\rangle$ and its counterpart 
$\vert\Psi_{\rm BG}\rangle$ do give the exact ground state
$\vert\Psi^{(4)}\rangle$ of the half-filled Hubbard square, but Kaplan's
generalization \cite{Kapl1} fails in doing so, because what should be
suppressed by his additional operator (configurations with doublons having no
holons on nearest-neighbour sites) is absent for the states
$\vert\Phi_D\rangle,\, D=0,1,2$. More involved variational states \cite{Taha1}
can easily reproduce the exact ground state.

The ground state $\vert\Psi^{(2)}\rangle$ for two electrons on the four sites
of a square is rather different. Its energy reaches the limiting value  
$E_\infty=-2\sqrt{2}\, t$ (and not 0) for $U\rightarrow\infty$. Its 
quasimomentum is 0 and the two particles form an $s$-wave pair. The quantity 
$\langle\Psi^{(2)}\vert C_s C^\dag_s\vert\Psi^{(2)}\rangle$, expressing the
affinity for $s$-wave pairing, dominates now over the
corresponding $d$-wave expression, 
$\langle\Psi^{(2)}\vert C_d C^\dag_d\vert\Psi^{(2)}\rangle$, for all values of 
$U$ except for $U=0$, where the two quantities are equal. The difference 
increases with $U$ and reaches a maximum for $U\rightarrow\infty$.

$\vert\Psi^{(2)}\rangle$ can also be represented as a superposition of 
three components, but now only one of them contains real-space configurations 
with doubly occupied sites. As a result, the Gutzwiller ansatz 
$\vert\Psi_{\rm G}\rangle$ gives a fairly poor description of the state, it
even predicts a wrong asymptotic value for the energy, 
$E_{\rm G}-E_\infty\approx0.16\, t$ for $U\rightarrow\infty$ (\fref{fig:DeltaE/E}(b)). The refined
Gutzwiller ansatz $\vert\Psi_{\rm GB}\rangle$ is not perfect either, it
reproduces the exact ground state only below a limiting value of $U\approx
3.46\, t$. Even if one adds a third operator $\rme^{-g'\hat{D}}$ in front of 
$\vert\Psi_{\rm GB}\rangle$, the situation remains essentially the same. What
needs to be done in order to achieve full coincidence with the exact ground
state for all $U$ is to use  $\hat P_{0}\hat H_0\hat P_{0}$ 
instead of $\hat H_0$ in combination with the Gutzwiller factor, thereby 
ensuring that no more doublons are created by $\hat H_0$ after being
suppressed by $\rme^{-g\hat D}$.

On the other hand, $\vert\Psi_{\rm B}\rangle$ gives a
more or less satisfactory approximation of the exact ground state, about as 
good as the Gutzwiller ansatz in the half-filled case. Adding to
$\vert\Psi_{\rm B}\rangle$ any of the above-mentioned factors (Gutzwiller,
doublon-holon correlations, Jastrow) gives full coincidence with the
exact two-electron ground state.

The analysis of the Hubbard square confirms that proceeding from
$\vert\Psi_{\rm G}\rangle$ to $\vert\Psi_{\rm GB}\rangle$ (or from
$\vert\Psi_{\rm B}\rangle$ to $\vert\Psi_{\rm BG}\rangle$) improves
substantially the variational ground state. Moreover, for $U\gtrsim t$ it is
preferable to use $\vert\Psi_{\rm B}\rangle$ ($\vert\Psi_{\rm BG}\rangle$) as a
variational ansatz, rather than $\vert\Psi_{\rm G}\rangle$ ($\vert\Psi_{\rm
  GB}\rangle$), {\it i.e.}\ it is advantageous to start from 
$\vert\Psi_\infty\rangle$, the ground state of the $t-J$ model in the limit
$J\rightarrow 0$ (or of the Heisenberg Hamiltonian at half filling).

\section{Antiferromagnetism}
\label{sec:AF}

We return now to the square lattice. The lessons from the Hubbard square
suggest the  use of $\vert\Psi_{\rm B}\rangle$ or $\vert\Psi_{\rm BG}\rangle$, 
but, unfortunately, the reference state $\vert\Psi_\infty\rangle$ is very
difficult to handle, except for small clusters (using exact
diagonalization). This is the main reason why $\vert\Psi_{\rm B}\rangle$ and
$\vert\Psi_{\rm BG}\rangle$ have not been used so far for the square
lattice. In the remainder of this paper we will therefore mostly be concerned
with wave functions linked to $\vert\Psi_0\rangle$, the ground state of an
appropriate mean-field Hamiltonian.

We first consider an antiferromagnetic ground state, restricting ourselves
to the half-filled band case (number of particles $N$ equal to the number of 
sites $L$). To enforce a commensurate spin-density wave (with a preference 
for up spins on one sublattice and for down spins on the other), we introduce 
the mean-field Hamiltonian
\begin{equation}
\hat{H}_{\rm mf}=\hat{H}_0-\Delta\sum_i(-1)^i(n_{i\uparrow}-n_{i\downarrow})\, ,
\end{equation}
where $i$ is even on one sublattice and odd on the other. The staggered order
parameter $m$ is defined as
\begin{equation}
m:=\frac{1}{2L}\sum_i(-1)^i\langle n_{i\uparrow} -n_{i\downarrow}\rangle\, ,
\end{equation} 
where the expectation value is calculated with respect to the chosen trial
state. The simplest case is the ground state of $\hat{H}_{\rm mf}$,
a commensurate spin-density wave (SDW) with wave vector
${\bi Q}=(\pi,\pi)$.

The expectation value of the Hubbard Hamiltonian can be written as
\begin{equation}
\langle\hat{H}\rangle=\langle\hat{H}_{\rm mf}\rangle+2L\Delta m
+U\sum_i\langle n_{i\uparrow}n_{i\downarrow}\rangle\, .
\end{equation}
For the trial state $\vert{\rm SDW}\rangle$, the ground state of 
$\hat{H}_{\rm mf}$, the average double occupancy is simply given by
\begin{equation}
\label{eq:do-mf}
\langle{\rm SDW}\vert n_{i\uparrow}n_{i\downarrow}\vert{\rm SDW}\rangle
=\frac{1}{4}-m^2\, ,
\end{equation}
while the Hellman-Feynman theorem yields
\begin{equation}
\label{eq:hf-mf}
\frac{\rmd}{\rmd\Delta}\langle{\rm SDW}\vert\hat{H}_{\rm
  mf}\vert{\rm SDW}\rangle=-2Lm\, .
\end{equation}
Minimizing $\langle\hat{H}\rangle$ with respect to $\Delta$ and using (\ref{eq:do-mf}) and (\ref{eq:hf-mf}), we obtain
\begin{equation}
(\Delta -Um)\, \frac{\rmd m}{\rmd\Delta}=0\, ,
\end{equation}
and therefore 
\begin{equation}
\label{eq:min-mf}
\Delta=Um\, .
\end{equation}

The order parameter $m$ is easily calculated for the mean-field state. 
The transformation 
\begin{equation}
c_{\bi k\sigma}=\frac{1}{\sqrt{L}}\sum_i\rme^{-\rmi{\bi k\, \cdot\bi R}_i}\, 
c_{i\sigma}\, ,
\end{equation}
where the wave vectors ${\bi k}$ belong to the first Brillouin zone and
${\bi R}_i,\, i=1...,L,$ are lattice vectors (with lattice constant set to $1$),
diagonalizes the hopping term, 
\begin{equation}
\hat{H}_0=\sum_{\bi k\sigma}\varepsilon_{\bi k}\, c_{\bi k\sigma}^\dag
c_{\bi k\sigma}\, ,
\end{equation}
with a spectrum
\begin{equation}
\varepsilon_{\bi k}=-2t(\cos k_x+\cos k_y)-4t'\cos k_x\cos k_y\, .
\end{equation}
In the rest of this section, we choose $t'=0$ and thus obtain a spectrum with 
electron-hole symmetry,
\begin{equation}
\varepsilon_{\bi k}=-\varepsilon_{\bi k\pm\bi Q}\, .
\end{equation}
In reciprocal space, the mean-field Hamiltonian reads
\begin{eqnarray}
\fl\hat{H}_{\rm mf}=\sum_{\bi k\sigma}\! '
\left\{\varepsilon_{\bi k}(c_{\bi k\sigma}^\dag c_{\bi k\sigma}
-c_{\bi k+\bi Q,\sigma}^\dag c_{\bi k+\bi Q,\sigma})\right.\left.-\sigma\Delta(c_{\bi k\sigma}^\dag c_{\bi k+\bi Q,\sigma}+c_{\bi k+\bi Q,\sigma}^\dag c_{\bi k\sigma})\right\}\, ,
\end{eqnarray}
where the notation $\sum_{\bi k}'$ means that the sum includes only half of the
wave vectors of the Brillouin zone, those satisfying 
$\varepsilon_{\bi k}\le 0$.
$\hat{H}_{\rm mf}$ is easily diagonalized by a Bogoliubov transformation. The
spectrum (in the folded zone) is split into $\pm E_{\bi k}$, where
\begin{equation}
E_{\bi k}=\sqrt{\varepsilon_{\bi k}^2+\Delta^2}\, .
\end{equation}
The order parameter is found to be
\begin{equation}
\label{eq:op-mf}
m=\frac{1}{L}\sum_{\bi k}\! '\, \frac{\Delta}{E_{\bi k}}\, .
\end{equation}
The relations (\ref{eq:min-mf}) and (\ref{eq:op-mf}) determine
both the gap parameter $\Delta$ and the order parameter $m$ for the
mean-field ground state $\vert{\rm SDW}\rangle$.

We now proceed to the correlated trial state
\begin{equation}
\label{eq:GB-SDW}
\vert\Psi_{\rm GB}\rangle
=\rme^{-h\hat{H}_0/t}\, \rme^{-g\hat{D}}\, \vert{\rm SDW}\rangle\, .
\end{equation}
To decouple the terms $n_{i\uparrow}n_{i\downarrow}$ in the operator 
$\rme^{-g\hat{D}}$, 
a discrete Hubbard-Stratonovich transformation \cite{Hirs2} is applied,
\begin{eqnarray}
\label{eq:HS}
\rme^{-g\hat D}&={\rm exp}\left\{-g\sum_{i} n_{i\uparrow}n_{i\downarrow}\right\}\nonumber\\ &
=\prod_{i}\cosh[2a(n_{i\uparrow}-n_{i\downarrow})]\;
{\rm exp}\left\{-\frac{g}{2}(n_{i\uparrow}+n_{i\downarrow})\right\}\nonumber\\ 
&=2^{-L}\sum_{\tau_1,...,\tau_L}
{\rm exp}\left\{\sum_{i\sigma}(2a\sigma\tau_{i}-\frac{g}{2})n_{i\sigma}\right\}\, ,
\end{eqnarray}
where $\tau_i$ are Ising variables assuming
the values $\pm 1$, and $a={\rm arctg}\sqrt{{\rm th}(g/4)}$. As a 
result, both exponential operators in (\ref{eq:GB-SDW}) are 
quadratic in fermionic creation and annihilation operators, and therefore the 
fermionic degrees of freedom can be integrated out \cite{Eich4}. We obtain the 
following expressions
\begin{eqnarray}
\langle\Psi_{\rm GB}\vert\hat{H}\vert\Psi_{\rm GB}\rangle
=\sum_{{\tau_1,...,\tau_L}\atop{\tau_1',...,\tau_L'}}
E(\tau_1,...\tau_L;\tau_1',...,\tau_L')\, ,\nonumber\\
\langle\Psi_{\rm GB}\vert\Psi_{\rm GB}\rangle
=\sum_{{\tau_1,...,\tau_L}\atop{\tau_1',...,\tau_L'}}
N(\tau_1,...\tau_L;\tau_1',...,\tau_L')\, ,
\end{eqnarray}
where the quantities $E(\tau_1,...\tau_L;\tau_1',...,\tau_L')$ and
$N(\tau_1,...\tau_L;\tau_1',...,\tau_L')$ are products of determinants of
$L/2\times L/2$ matrices. (Initially, one has to deal with $L\times L$ 
matrices, but these can be block-diagonalized into two $L/2\times L/2$ matrices
because up and down spins are never mixed.)

The expectation value of the energy
\begin{equation}
E(g,h,\Delta)=\frac{\langle\Psi_{\rm GB}\vert\hat{H}\vert\Psi_{\rm GB}\rangle}
{\langle\Psi_{\rm GB}\vert\Psi_{\rm GB}\rangle}
\end{equation}
is then calculated by Monte Carlo sampling and minimized with respect to
the three variational parameters $g,h,\Delta$. The computations have been
limited to an $8\times 8$ lattice, with a fixed value of $U=8\, t$.

It is instructive to follow the evolution of the results as the 
variational ground state is refined, from the simple spin-density wave state
$\vert\Psi_0\rangle=\vert{\rm SDW}\rangle$ via the Gutzwiller ansatz
$\vert\Psi_{\rm G}\rangle=\rme^{-g\hat{D}}\vert{\rm SDW}\rangle$ to
$\vert\Psi_{\rm
  GB}\rangle=\rme^{-h\hat{H}_0/t}\rme^{-g\hat{D}}\vert{\rm SDW}\rangle$. The
gap parameter $\Delta$ is found to decrease dramatically \cite{Eich2}. It
amounts to about $3.6\, t$ for the plain SDW state, $1.3\, t$ for the
Gutzwiller ansatz and $0.32\, t$ for $\vert\Psi_{\rm GB}\rangle$. This indicates
that $\Delta$ has no clear physical meaning and cannot be simply related to an
excitation gap for the correlated states $\vert\Psi_{\rm G}\rangle$ and
$\vert\Psi_{\rm GB}\rangle$. We expect $\Delta$ to decrease steadily as the 
variational ansatz is further refined and to tend to zero as the trial state 
approaches the exact ground state. In fact, in this limit the gap parameter
merely plays the role of an (infinitesimal) symmetry-breaking field.

In spite of the strong variation of $\Delta$, 
the order parameter does not change much \cite{Eich2}, from $m\approx 0.45$ 
for $\vert{\rm SDW}\rangle$ to $m\approx 0.43$ for $\vert\Psi_{\rm G}\rangle$
and $m\approx 0.39$ for $\vert\Psi_{\rm GB}\rangle$. Thus the result for our
most sophisticated variational ansatz is higher than the extrapolation
$m\approx 0.22$ from Monte Carlo simulations on relatively small
lattices \cite{Hirs4}, and also higher than the value $m\approx 0.31$
extracted from large-scale Monte Carlo simulations \cite{Sand1} for the 2D
spin-$\frac{1}{2}$ Heisenberg model (and expected to represent an upper limit
for the antiferromagnetic order parameter of the Hubbard model).

Quite generally, the order parameter obtained within mean-field theory is
reduced when quantum fluctuations are taken into account. Our results indicate
that quantum fluctuations are progressively included when proceeding from the
simple spin-density wave state $\vert{\rm SDW}\rangle$ over $\vert\Psi_{\rm
  G}\rangle$ to $\vert\Psi_{\rm GB}\rangle$. Nonetheless, long-range (spin
wave) fluctuations are suppressed by the gap $\Delta$, which is small but
still finite for $\vert\Psi_{\rm GB}\rangle$. It is interesting to note that a
recent calculation using a Quantum Cluster method \cite{Kanc1} gave
$m\approx0.4$, which agrees with our value. The small size of the cluster
($2\times 2$) used in these calculations suggests that long-range fluctuations
are not properly taken into account either.

\section{Superconductivity}
\label{sec:SC}

We discuss now the region away from half filling, $\it i.e.$\,
densities $n=N/L\neq 1$. We have tried to examine the fate of 
antiferromagnetism for $n\neq 1$, but so far our variational Monte Carlo 
approach did not converge fast enough to allow the extraction of reliable 
results. In the following we limit ourselves to superconducting ground 
states with $d$-wave symmetry. The mean-field state $\vert\Psi_0\rangle$
is constructed as a conventional Bardeen-Cooper-Schrieffer (BCS) state, with
\begin{equation}
\Delta({\bi k})=\Delta_0(\cos k_x-\cos k_y)\, .
\end{equation}
The variational calculations follow the same steps as in the SDW case,
but now one has to deal with $2L\times 2L$ matrices because, in contrast to 
the SDW case, up and down spins are now mixed as are particles and holes.

Our trial state 
\begin{equation}
\vert\Psi_{\rm GB}\rangle=\rme^{-h\hat{H}_0/t}\rme^{-g\hat{D}}\vert{\rm BCS}\rangle
\label{eq:GB-BCS}
\end{equation}
has three variational parameters, $h,g$ and $\Delta_0$, as well as the 
parameter $\mu$, which fixes the average number of 
electrons, but is not identical to the true chemical potential. Instead of
this `grand-canonical' set-up one could also work with a BCS state projected
onto a fixed number of particles, where $\mu$ becomes a fourth variational
parameter. Unfortunately, the minus sign problem turns out to be severe in the 
`canonical' case \cite{Eich1}, presumably because 
$\vert{\rm BCS}\rangle_N$ is a correlated state, whereas the conventional 
BCS wave function can be written as a single Slater determinant. The results 
discussed below have all been obtained using the grand-canonical version.

The calculations were mostly done for an $8\times8$ lattice, but in order to 
study the size dependence we have also made a few runs for $6\times6$ and
$10\times 10$ lattices. The results for an $8\times 8$ and a
$10\times 10$ lattice do not differ considerably if the gap parameter 
$\Delta_0$ is large enough \cite{Eich2}, as is the case for $n=0.90$ and 
$0.94$. For $n=0.84$ the finite size effects are still rather strong.

As in the case of antiferromagnetism, the gap parameter 
$\Delta_0$ varies strongly as the trial wave function is refined. Thus, the 
fully Gutzwiller-projected BCS state, used as an ansatz for the $t-J$ 
model \cite{Para1}, gives large values in an extended region of densities.
Within this approximation, the largest gap parameter, $\Delta_0\approx t$, 
is found at half filling, followed by a nearly linear decrease as a 
function of doping concentration $x=1-n$, until $\Delta_0$ vanishes at 
$x_{\rm c}\approx0.35$. Similar behaviour is found for $\vert\Psi_{\rm G}\rangle$
applied to the Hubbard model \cite{Eich2}, but with a smaller critical
density and a reduced value at $x=0$, $\Delta_0\approx 0.2\, t$. The gap
is reduced still further when proceeding from $\vert\Psi_{\rm G}\rangle$ to
$\vert\Psi_{\rm GB}\rangle$ \cite{Eich2}, with a maximum of $0.13\, t$ for
$x\approx 0.1$ and a critical hole density $x_{\rm c}\approx 0.18$. It is
worthwhile to mention that a Gutzwiller ansatz with coexisting
antiferromagnetism and superconductivity also produces a maximum in the gap
parameter as a function of density \cite{Giam1}, $\Delta_0\approx0.1\, t$ at
$x\approx 0.1$.

For the repulsive Hubbard model, the superconducting order parameter $\Phi$
is commonly chosen as the expectation value of a pair of creation operators on
neighbouring sites, 
\begin{equation}
\label{eq:op-supra}
\Phi=\langle c_{i\uparrow}^\dag c_{i+\tau\downarrow}^\dag\rangle\, .
\end{equation}
For $d$-wave symmetry, $\Phi$ has a different sign for a horizontal 
bond than for a vertical bond. Alternatively, one can extract the order 
parameter from the correlation function 
\begin{equation}
\label{eq:cf-supra}
S_{\tau\tau'}({\bi R}_i-{\bi R}_j)=\langle c_{i+\tau\downarrow}c_{i\uparrow}
c_{j\uparrow}^\dag c_{j+\tau'\downarrow}^\dag\rangle
\end{equation} 
for $\vert {\bi R}_i-{\bi R}_j\vert\rightarrow\infty$. For the exact ground 
state, the two procedures, the evaluation of (\ref{eq:op-supra}) for an infinitesimal symmetry breaking on
the one hand, and the determination of the asymptotic behaviour of the
correlation function (\ref{eq:cf-supra}) on the other, are expected to 
yield the same order parameter $\Phi$ in the thermodynamic limit 
($N,L\rightarrow\infty\mbox{ for constant density }n$).

\begin{figure}
\centering
\includegraphics[width=0.6\textwidth]{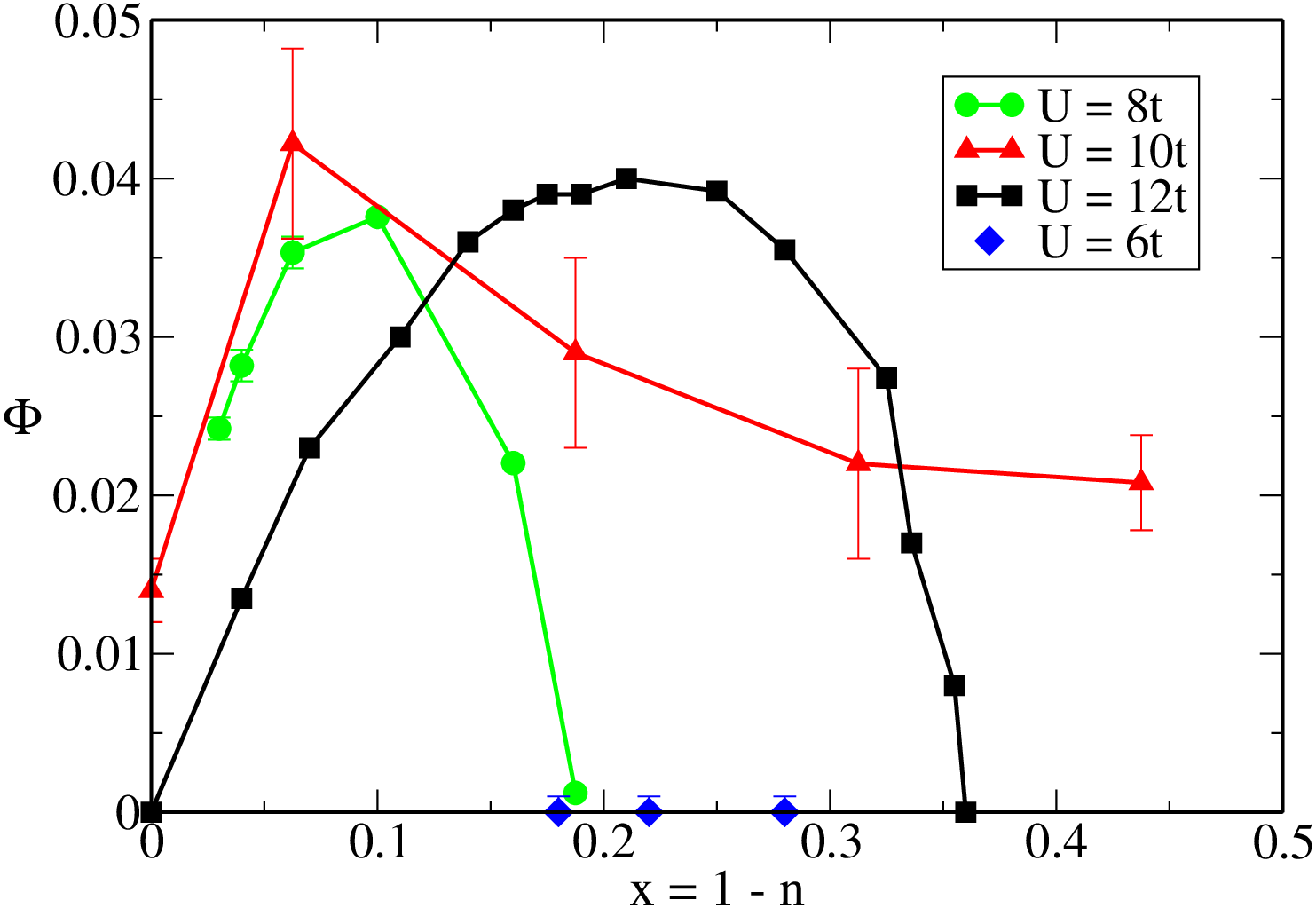}
\caption{Comparison of superconducting order parameters for the simple Hubbard 
model and four different approaches: full Gutzwiller projection for the $t-J$
model ($U=12\, t,\, \fullsquare$---$\fullsquare$), partial Gutzwiller
projection ($U=10\, t,\,
{\color{red}\fulltriangle}\!${\color{red}---}$\!{\color{red}\fulltriangle}$),
our ansatz ($U=8\, t,\,
{\color{green}\fullcircle}\!\!\!${\color{green}---}$\!{\color{green}\fullcircle}$),
and Gaussian Monte Carlo ($U=6\, t,\,
{\color{blue}\fulldiamond}$).}
\label{fig:op1}
\end{figure}

We first discuss results for $t'=0$. \Fref{fig:op1} shows the 
$d$-wave order parameter as a function of doping, $x=1-n$, obtained following
four different routes. Three of them are variational, full 
Gutzwiller projection for the $t-J$ model (black squares) \cite{Para1}, 
the Gutzwiller ansatz, including a finite antiferromagnetic order parameter 
close to half filling (red triangles) \cite{Giam1}, and our ansatz 
(\ref{eq:GB-BCS}) (green circles) \cite{Eich2}. The null result 
(blue diamonds) is from recent (Gaussian) Monte Carlo simulations of Aimi and
Imada \cite{Aimi1}, who extracted tiny upper bounds for the order parameter
from the long-distance behaviour of the correlation function 
(\ref{eq:cf-supra}). Unfortunately, the
four data sets were obtained for four different values of $U$.
Nonetheless, the three variational results are remarkably similar for weak
doping ($x \lesssim 0.1$). For $x\gtrsim 0.18$ there appears to be a
discrepancy between the results of \cite{Giam1} and our results. Note that the
Gutzwiller data \cite{Giam1} have been obtained on the basis of the
correlation function (\ref{eq:cf-supra}), which is expected to yield a finite order parameter for a finite
system, even if there is no long-range order in the thermodynamic
limit. Therefore the apparent discrepancy may be an artefact of the procedure,
as pointed out in \cite{Giam1}. This is confirmed by our own results for the
Gutzwiller trial state $\vert\Psi_{\rm G}\rangle$, where we find \cite{Eich4}
that the extrapolation of the order parameter for $L\rightarrow\infty$ gives
$\Phi\rightarrow0$ for $x\gtrsim0.2$, in agreement with our results for
$\vert\Psi_{\rm GB}\rangle$. Clearly our ansatz $\vert\Psi_{\rm GB}\rangle$
should be superior to the Gutzwiller wave function, either partially or fully
projected. It is worthwhile to mention that refined variational states for the
$t-J$ model \cite{Sore1} did not give markedly different results from
those presented in \fref{fig:op1} for the fully Gutzwiller-projected
wave function. Therefore there seems to be a clear difference between 
the Hubbard and $t-J$ model predictions for superconductivity above, say,
a doping concentration of 18\%, where superconductivity appears to be absent
for the Hubbard model, but to persist up to about 40\% for the $t-J$ model. 
This is not a true discrepancy because the mapping from the large-$U$ Hubbard 
model to the $t-J$ model is only justified close to half filling.
\Fref{fig:op1} also shows that there is no disagreement between our
variational result \cite{Eich2} and the Gaussian Monte Carlo data
\cite{Aimi1}, which so far are only available for $x\gtrsim 0.18$.

\begin{figure}
\centering
\includegraphics[width=0.6\textwidth]{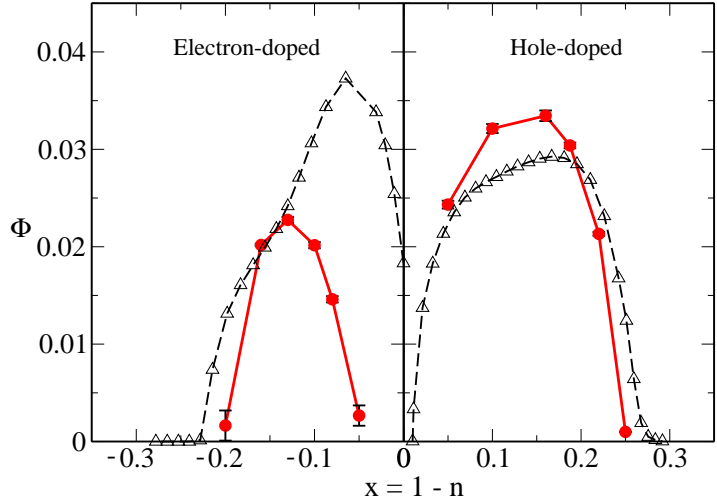}
\caption{Superconducting order parameter as a function of doping for the
  Hubbard model including next-nearest-neighbour hopping:
  ${\color{red}\fullcircle}\!\!\!${\color{red}---}$\!{\color{red}\fullcircle}$\; our variational data, $\opentriangle\!$---$\!\opentriangle$\;results obtained with the Quantum Cluster method. All data points were obtained using the parameter values\, $U=8\, t,\; t'=-0.3\, t$.}
\label{fig:op2}
\end{figure}

We turn now to the more realistic case of hopping between both nearest and 
next-nearest neighbours. \Fref{fig:op2} compares our
variational results (red circles) \cite{Eich3} with recent data obtained 
with the Quantum Cluster method (open triangles) \cite{Kanc1}. The same 
parameter sets ($U=8\, t,\; t'=-0.3\, t$) were used in both approaches. The
agreement is remarkably good, in view of the fact that the two methods are
very different. Only the surprising peak found in the Quantum Cluster approach
for an electron doping of about 5\% is absent in our variational data. This
peak is suppressed if antiferromagnetic long-range order is taken into account \cite{Kanc1}. 

As noted previously \cite{Eich3}, the interval $-0.2\lesssim x\lesssim 0.25$, 
to which 
superconductivity is confined according to our variational study, corresponds 
to densities where the Fermi surface passes through the boundary of the 
folded Brillouin zone, at so-called hot spots. This means that parts of the 
Fermi surface are connected by wave vectors close to ${\bi Q}=(\pi,\pi)$, for 
which the magnetic structure factor has a pronounced peak \cite{Eich3}, 
reminiscent of long-range antiferromagnetic order at half filling. 
This observation lends support to a
magnetic pairing mechanism. In fact, an effective attraction of
the form $-(U^2/t)\, \chi_0({\bi k+\bi k'})$ is deduced using second-order
perturbation theory for the effective particle-particle vertex \cite{Binz1},
where $\chi_0({\bi q})$ is the spin susceptibility for non-interacting
electrons and the incoming and outgoing electrons have wave vectors 
${\pm\bi k}$ and ${\pm\bi k'}$, respectively. A similar expression for the
effective attraction, $-\frac{3}{2}(\overline{U}^2/t)\, \chi({\bi q},\omega)$,
has been fitted to the particle-particle vertex calculated with the
Quantum Cluster method \cite{Maie5}, with a density- and temperature-dependent 
coupling constant $\overline{U}$ and the numerically calculated spin 
susceptibility $\chi({\bi q},\omega)$.

\section{Summary and concluding remarks}
\label{sec:sum}

In this paper, we have analysed various variational wave functions in the 
context of the 2D Hubbard model. Both antiferromagnetic ground states (at
half filling) and superconducting ground states with $d$-wave symmetry
(away from half filling) have been studied. We have considered states of 
the general form
\begin{equation}
\label{eq:ansatz}
\vert\Psi\rangle=\hat{P}\vert\Psi_0\rangle\, ,
\end{equation}
where $\vert\Psi_0\rangle$ is the ground state of an appropriate mean-field 
Hamiltonian, which includes a variational parameter $\Delta$, the gap 
parameter. 
The operator $\hat{P}$ is unity in the simple mean-field theory, it consists of
a single operator, $\hat{P}=\rme^{-g\hat{D}}$, in the Gutzwiller ansatz or 
of several operators in more refined wave functions. The optimized gap
parameter was found to depend sensitively on the choice of the wave function.
It decreases by about an order of magnitude when proceeding from the simple
mean-field theory to our favourite ansatz with 
$\hat{P}=\rme^{-h\hat{H}_0/t}\rme^{-g\hat{D}}$. Therefore $\Delta$, taken
as a variational parameter in wave functions of the form 
(\ref{eq:ansatz}), cannot be identified with a physical energy gap
(except in the simple BCS case where $\Delta$ is the minimum energy for adding 
an electron to the system), nor can it be associated with a characteristic
energy scale such as $k_{\rm B}T^*$ (where $T^*$ is the so-called pseudogap
temperature \cite{Ande5, Para1}). In contrast, the order parameter was found 
to depend little
on the sophistication of the variational ansatz.

To our knowledge, there is no rigorous proof for the existence of long-range
antiferromagnetic order in the 2D Hubbard model at half filling (nor in the
case of the spin-$\frac{1}{2}$ Heisenberg model), although this question is
not strongly debated. The case of ($d$-wave) superconductivity is much more
controversial.
Both the relatively small disparities between the order parameters 
obtained with different variational states close to half filling and
the generally good agreement between our results and those of completely 
different approaches, Quantum Cluster \cite{Kanc1} and Gaussian Monte Carlo 
\cite{Aimi1}, give rather strong support for the existence of $d$-wave 
superconductivity in the Hubbard model on a square lattice. Nevertheless,
as we have seen in the simple example of the Hubbard square, our wave
function is probably not optimal for the large value of $U$ used in our 
calculations (and believed to be appropriate for the cuprates). Progress with
complementary wave functions (linked to the $U\rightarrow\infty$ limit) or
with other methods would be highly desirable.

\ack
We are grateful for financial support from the Swiss National Science 
Foundation through the National Centre of Competence in Research 
`Materials with Novel Electronic Properties-MaNEP'.


\section*{References}

\begin{thebibliography}{10}

\bibitem{Kohn1}
W. Kohn and J.M. Luttinger, Phys.\ Rev.\ Lett. {\bf 15},  524  (1965).

\bibitem{Ande1}
P.W. Anderson, Science {\bf 235},  1196  (1987).

\bibitem{Mish1}
A.S. Mishchenko, N. Nagaosa, Z.-X. Shen, G. De{ }Filippis, V. Cataudella, T.P.
  Devereaux, C. Bernhard, K.W. Kim, and J. Zaanen, Phys.\ Rev.\ Lett. {\bf 100},
  166401  (2008).

\bibitem{Gius1}
F. Giustino, M.L.\ Cohen, and S.G.\ Louie, Nature {\bf 452},  975  (2008).

\bibitem{Rezn1}
D. Reznik, G. Sangiovanni, O. Gunnarsson, and T.P. Devereaux, Nature {\bf 455},
  E6  (2008).

\bibitem{Binz2}
B. Binz, D. Baeriswyl, and B. Dou{\c{c}}ot, Ann.\ Phys.\ (Leipzig) {\bf 12},
  704  (2003).

\bibitem{Zanc1}
D. Zanchi and H.J. Schulz, Phys.\ Rev.\ B {\bf 61},  13609  (2000).

\bibitem{Halb1}
C.J. Halboth and W. Metzner, Phys.\ Rev.\ B {\bf 61},  7364  (2000).

\bibitem{Hone3}
C. Honerkamp and M. Salmhofer, Phys.\ Rev.\ B {\bf 64},  184516  (2001).

\bibitem{Binz1}
B. Binz, D. Baeriswyl, and B. Dou{\c{c}}ot, Eur.\ Phys.\ J.\ B {\bf 25},  69
  (2002).

\bibitem{Ande3}
P.W. Anderson, J.\ Phys.\ Chem.\ Solids {\bf 11},  26  (1959).

\bibitem{Chao1}
K.A. Chao, J. Spa\l{}ek, and A.M. Ole\'s, Phys.\ Rev.\ B {\bf 18},  3453  (1978).

\bibitem{Manu1}
E. Manousakis, Rev.\ Mod.\ Phys. {\bf 63},  1  (1991).

\bibitem{Lee1}
P.A. Lee, N. Nagaosa, and X.-G. Wen, Rev.\ Mod.\ Phys. {\bf 78},  17  (2006).

\bibitem{Ogat1}
M. Ogata and H. Fukuyama, Rep.\ Prog.\ Phys. {\bf 71},  036501  (2008).

\bibitem{Gros1}
C. Gros, Phys.\ Rev.\ B {\bf 38},  931  (1988).

\bibitem{Yoko1}
H. Yokoyama and H. Shiba, J.\ Phys.\ Soc.\ Jpn. {\bf 57},  2482  (1988).

\bibitem{Para1}
A. Paramekanti, M. Randeria, and N. Trivedi, Phys.\ Rev.\ B {\bf 70},  054504
   (2004).

\bibitem{Path1}
S. Pathak, V.B. Shenoy, M. Randeria, and N. Trivedi, Phys.\ Rev.\ Lett. {\bf
  102},  027002  (2009).

\bibitem{Paul1}
L. Pauling, {\em The Nature of the Chemical Bond} (Cornell University Press,
  Ithaca, N.Y., 1967).

\bibitem{Ande5}
P.W. Anderson, P.A. Lee, M. Randeria, T.M. Rice, N. Trivedi, and F.C. Zhang, J.\ Phys.\
  Condens.\ Matter {\bf 16},  R755  (2004).

\bibitem{Cold1}
R. Coldea, S.M. Hayden, G. Aeppli, T.G. Perring, C.D. Frost, T.E. Mason, S.-W. Cheong,
  and Z. Fisk, Phys.\ Rev.\ Lett. {\bf 86},  5377  (2001).

\bibitem{Kata1}
A.A. Katanin and A.P. Kampf, Phys.\ Rev.\ B {\bf 66},  100403  (2002).

\bibitem{Muel1}
E. M$\ddot{\rm u}$ller-Hartmann and A. Reischl, Eur.\ Phys.\ J.\ B {\bf 28},
  173  (2002).

\bibitem{Kata2}
A.A. Katanin and A.P. Kampf, Phys.\ Rev.\ B {\bf 67},  100404  (2003).

\bibitem{Coma1}
A. Comanac, L. de'{ }Medici, M. Capone, and A.J. Millis, Nature Phys. {\bf 4},
  287  (2008).

\bibitem{Aebi1}
P. Aebi, J. Osterwalder, P. Schwaller, L. Schlapbach, M. Shimoda, T. Mochiko,
  and K. Kadowaki, Phys.\ Rev.\ Lett. {\bf 72},  2757  (1994).

\bibitem{Bori1}
S.V. Borisenko, M.S. Golden, S. Legner, T. Pichler, C. D$\ddot{\rm u}$rr, M.
  Knupfer, J. Fink, G. Yang, S. Abell, and H. Berger, Phys.\ Rev.\ Lett. {\bf
  84},  4453  (2000).

\bibitem{Norm1}
M.R. Norman, M. Randeria, H. Ding, and J.C. Campuzano, Phys.\ Rev.\ B {\bf 52},
  615  (1995).

\bibitem{Kim1}
C. Kim, P.J. White, Z.-X. Shen, T. Tohyama, Y. Shibata, S. Maekawa, B.O. Wells, Y.J.
  Kim, R.J. Birgenau, and M.A. Kastner, Phys.\ Rev.\ Lett. {\bf 80},  4245  (1998).

\bibitem{Dago1}
E. Dagotto, Rev.\ Mod.\ Phys. {\bf 66},  763  (1994).

\bibitem{Troy1}
M. Troyer and U.-J. Wiese, Phys.\ Rev.\ Lett. {\bf 94},  170201  (2005).

\bibitem{Corn1}
J.F. Corney and P.D. Drummond, Phys.\ Rev.\ B {\bf 73},  125112  (2006).

\bibitem{Corb1}
B. Corboz, M. Troyer, A. Kleine, I.P. McCulloch, U. Schollw$\ddot{\rm o}$ck, and
  F.F. Assaad, Phys.\ Rev.\ B {\bf 77},  085108  (2008).

\bibitem{Geor1}
A. Georges, G. Kotliar, W. Krauth, and M.J. Rozenberg, Rev.\ Mod.\ Phys. {\bf
  68},  13  (1996).

\bibitem{Maie2}
T. Maier, M. Jarrell, T. Pruschke, and M.H. Hettler, Rev.\ Mod.\ Phys. {\bf 77},
  1027  (2005).

\bibitem{Sene1}
D. S\'en\'echal, {\em An introduction to quantum cluster methods},
arXiv:0806.2690.

\bibitem{Park1}
H. Park, K. Haule, and G. Kotliar, Phys.\ Rev.\ Lett. {\bf 101},  186403
  (2008).

\bibitem{Gutz1}
M.C. Gutzwiller, Phys.\ Rev.\ Lett. {\bf 10},  159  (1963).

\bibitem{Copp1}
S.N. Coppersmith and C.C. Yu, Phys.\ Rev.\ B {\bf 39},  11464  (1989).

\bibitem{Yoko4}
H. Yokoyama and H. Shiba, J.\ Phys.\ Soc.\ Jpn. {\bf 59},  3669  (1990).

\bibitem{Kapl1}
T.A. Kaplan, P. Horsch, and P. Fulde, Phys.\ Rev.\ Lett. {\bf 49},  889  (1982).

\bibitem{Yoko2}
H. Yokoyama, Y. Tanaka, M. Ogata, and H. Tsuchiura, J.\ Phys.\ Soc.\ Jpn. {\bf
  73},  1119  (2004).

\bibitem{Tocc1}
L.F. Tocchio, F. Becca, A. Parola, and S. Sorella, Phys.\ Rev.\ B {\bf 78},
  041101(R)  (2008).

\bibitem{Taha1}
D. Tahara and M. Imada, J.\ Phys.\ Soc.\ Jpn. {\bf 77},  114701  (2008).

\bibitem{Eich2}
D. Eichenberger and D. Baeriswyl, Phys.\ Rev.\ B {\bf 76},  180504(R)  (2007).

\bibitem{Eich3}
D. Eichenberger and D. Baeriswyl, Phys.\ Rev.\ B {\bf 79},  100510(R)  (2009).

\bibitem{Otsu1}
H. Otsuka, J.\ Phys.\ Soc.\ Jpn. {\bf 61},  1645  (1992).

\bibitem{Baer3}
D. Baeriswyl, Springer Series in Solid-State Sciences {\bf 69},  183  (1987).

\bibitem{Sore1}
S. Sorella, G.B. Martins, F. Becca, C. Gazza, L. Capriotti, A. Parola, and E.
  Dagotto, Phys.\ Rev.\ Lett. {\bf 88},  117002  (2002).

\bibitem{Dzie1}
M. Dzierzawa, D. Baeriswyl, and M. Di{ }Stasio, Phys.\ Rev.\ B {\bf 51},  1993
  (1995).

\bibitem{Dzie2}
M. Dzierzawa, D. Baeriswyl, and L.M. Martelo, Helv.\ Phys.\ Acta {\bf 70},  124
  (1997).

\bibitem{Mart1}
L.M. Martelo, M. Dzierzawa, L. Siffert, and D. Baeriswyl, Z.\ Phys.\ B {\bf 103},
   335  (1997).

\bibitem{Baer4}
D. Baeriswyl, Found.\ Phys. {\bf 30},  2033  (2000).

\bibitem{Lieb1}
E.H. Lieb, Phys.\ Rev.\ Lett. {\bf 62},  1201  (1989).

\bibitem{Scal3}
D.J. Scalapino and S.A. Trugman, Philos.\ Mag.\ B {\bf 74},  607  (1996).

\bibitem{Hirs2}
J.E. Hirsch, Phys.\ Rev.\ B {\bf 28},  4059  (1983).

\bibitem{Eich4}
D. Eichenberger, {\em Superconductivity and antiferromagnetism in the
  two-dimensional Hubbard model}, {P}h.\ {D}.\ thesis, {U}niversity of {F}ribourg, 2008 (unpublished).

\bibitem{Hirs4}
J.E. Hirsch and S. Tang, Phys.\ Rev.\ Lett. {\bf 62},  591  (1989).

\bibitem{Sand1}
A.W. Sandvik, Phys.\ Rev.\ B {\bf 56},  11678  (1997).

\bibitem{Kanc1}
S.S. Kancharla, B. Kyung, D. S\'en\'echal, M. Civelli, M. Capone, G. Kotliar, and
  A.-M. Tremblay, Phys.\ Rev.\ B {\bf 77},  184516  (2008).

\bibitem{Eich1}
D. Eichenberger and D. Baeriswyl, Physica C {\bf 460},  1153  (2007).

\bibitem{Giam1}
T. Giamarchi and C. Lhuillier, Phys.\ Rev.\ B {\bf 43},  12943  (1991).

\bibitem{Aimi1}
T. Aimi and M. Imada, J.\ Phys.\ Soc.\ Jpn. {\bf 76},  113708  (2007).

\bibitem{Maie5}
T.A. Maier, A. Macridin, M. Jarrell, and D.J. Scalapino, Phys.\ Rev.\ B {\bf 76},
  144516  (2007).

\end{thebibliography}
\end{document}